\documentclass[twocolumn,english,showpacs,preprintnumbers,amsmath,amssymb,aps]{revtex4}
\usepackage[T1]{fontenc}
\usepackage[latin9]{inputenc}
\usepackage{units}
\usepackage{graphicx}
\usepackage{amssymb}

\makeatletter
\@ifundefined{textcolor}{}
{ \definecolor{BLACK}{gray}{0}
 \definecolor{WHITE}{gray}{1}
 \definecolor{RED}{rgb}{1,0,0}
 \definecolor{GREEN}{rgb}{0,1,0}
 \definecolor{BLUE}{rgb}{0,0,1}
 \definecolor{CYAN}{cmyk}{1,0,0,0}
 \definecolor{MAGENTA}{cmyk}{0,1,0,0}
 \definecolor{YELLOW}{cmyk}{0,0,1,0}
 }

\makeatother

\usepackage{babel}

\begin{document}

\title{Fluctuation theorem in driven nonthermal systems with quenched disorder}
\author{J.A. Drocco$^1$, C.J. Olson Reichhardt$^2$, 
and C. Reichhardt$^2$}
\affiliation{
$^1$Department of Physics, Princeton University, Princeton, NJ 08544\\ 
$^2$Theoretical Division, Los Alamos National Laboratory, Los Alamos,
NM 87545}

\date{\today}
\begin{abstract}
We demonstrate that the fluctuation theorem of
Gallavotti and Cohen can be
used to characterize the class of dynamics that
arises in  nonthermal systems of collectively interacting   
particles driven over random quenched disorder. 
By observing the frequency
of entropy-destroying trajectories, we show that there are specific
dynamical regimes near depinning in which this theorem holds. 
Hence the 
fluctuation theorem 
can be used to characterize a significantly wider class of 
non-equilibrium systems than previously considered. 
We discuss how the 
fluctuation theorem 
could be tested in
specific systems where noisy dynamics appear at the transition from a pinned to
a moving phase 
such as in vortices in type-II superconductors,
magnetic domain walls, and dislocation dynamics. 
\end{abstract}


\maketitle

\vskip2pc 

The 
fluctuation theorem (FT) of 
Gallavotti and Cohen has been described as
a generalization of the second law of thermodynamics in systems outside
the thermodynamic limit \citep{Gallavotti,EvansCohen,EvansSearles}. It relates
the frequency of entropy-destroying, also sometimes called second-law-violating,
trajectories to entropy-creating trajectories and is sufficiently
general to apply to systems far from equilibrium. It has been demonstrated
to hold analytically for a class of time-reversible dynamical systems
\citep{Gallavotti}, and has been verified numerically in many others
\citep{SchmickMarkus,Aumaitre}. Wang \textit{et al.} provided the
first experimental verification of the relation by observing the fluctuations
of a dielectric particle pulled by an optical trap \citep{WangSevick}.
It has also been shown that the FT holds in some driven 
non-thermal systems such as granular materials \citep{Feitosa}
and a ball moving in a Sinai
billiard potential \citep{SchmickLiu}.

The 
FT has not previously been examined in the class of
nonthermal nonequilibrium systems
consisting of collectively interacting
particles  
moving over a random background, where 
noisy dynamics can occur near pinned to moving transitions.
Examples of this type of dynamics include the motion of 
magnetic domain walls
\cite{MW}, 
depinning of electron crystals in  solid-state materials \citep{Wig,Cooper}, 
plastic deformations in driven superconducting vortex
matter \citep{Vortex1,Vinokur,Vortex2,Higgins}, 
and colloidal particles moving over
quenched disorder \citep{Westervelt,Ling}. Closely related to these systems
are the dynamics of interacting dislocations under a strain \cite{Miguel}.  
Typically, there is a regime near the onset of motion 
where the particle trajectories are strongly disordered and    
$1/f$ or crackling noise arises. 
In this work, we show that the 
FT can be used to 
characterize the dynamical behavior of a general model 
of this type,
indicating that the 
FT could be applied to a much
wider range of nonequilibrium systems 
than previously considered and 
may hold in the general class of systems exhibiting crackling noise.  

We specifically examine the formulation of the 
FT given in Ref.~\citealp{SchmickMarkus}.
One of the main predictions of the FT
is that the probability density function (PDF) of the injected power $p(J_{\tau})$
obeys the following relation:
\begin{equation}
\frac{p(J_{\tau})}{p(-J_{\tau})}  =e^{J_{\tau}S_{\tau}}\label{eq:Schmickfluct} ,
\end{equation}
where $J_{\tau}$is the injected power, $\tau$ is the duration of
the trajectory, and $S_{\tau}$ is some constant. If $S_{\tau}$ varies
such that 
$\beta_{\tau}=\tau/S_{\tau}$ 
is constant for
$\tau\gg\Gamma_{i}$, where $\Gamma_{i}$ are the microscopic time
scales of the system, then we say that $\beta_{\infty}$ represents
an ``effective temperature.''  
Wang {\it et al.} \cite{WangSevick} experimentally measured the quantity in
Eq.~1 from 
the trajectories of a colloid that 
was driven through a thermal system. 
In the system we consider, there is no thermal bath; instead, the
particle trajectories are generated in the presence of an external drive,
a random quenched background, and interactions with other particles. 

We consider colloidal spheres confined to two dimensions and
driven with an electric field 
in the presence of randomly distributed pinning sites.
This particular model system has been shown to exhibit
the same general dynamical features, 
including plastic flow and moving crystalline phases \cite{Reichhardt},
observed in other collectively interacting particle 
systems driven over random disorder such
as vortices in type-II superconductors \cite{Vortex1,Vortex2}; thus we believe
the behavior in our system will be generic to other systems of this type. 
Additionally,
experimental realizations of this system permit the direct measurement of the
particle trajectories \cite{Ling}. 
We simulate a system of $N_c$ colloids with 
periodic boundary conditions in the $x$ and $y$ directions, and employ 
overdamped dynamics such that
the equation of motion for a single colloid $i$ is \begin{equation}
\eta\frac{d{\bf r}_{i}}{dt}={\bf f}_{T}^{i}+{\bf f}_{Y}^{i}+{\bf f}_{p}^{i}+
{\bf f}_{d}\end{equation}
 All quantities are rescaled to dimensionless units, and the damping
constant $\eta$ is set to unity. The thermal force ${\bf f}_{T}^{i}$ arises
from random Langevin kicks with the properties 
$\langle{\bf f}_{T}^{i}\rangle=0$
and $\langle{\bf f}_{T}^{i}(t){\bf f}_{T}^{j}(t^{\prime})\rangle=2\eta k_{B}T\delta(t-t^{\prime})\delta_{ij}$.
The colloid interaction force ${\bf f}_{Y}^{i}$ is given by the following
screened Coulomb repulsion: 
${\bf f}_{Y}^{i}=\sum_{j\ne i}^{N_{c}}A_{c}(\frac{4}{r_{ij}}+\frac{1}{r_{ij}^{2}})e^{-4r_{ij}}{\bf \hat{r}}_{ij}$.
Here $A_{c}$ is an adjustable coefficient, 
${\bf r}_{i(j)}$ is the position of vortex $i(j)$,
$r_{ij}=|{\bf r}_{i}-{\bf r}_j|$ 
and ${\bf {\hat r}}_{ij}=({\bf r}_i-{\bf r}_j)/r_{ij}$.
The quenched disorder introduces a force
${\bf f}_{p}^{i}$ which is modeled by $N_p$ randomly placed 
attractive parabolic pinning sites of strength $A_p$ and radius $r_p=0.45$,
${\bf f}_{p}^{i}=\sum_{k=1}^{N_{p}}(-A_{p}r_{ik}/r_p)\Theta(r_{p}-r_{ik}){\bf \hat{r}}_{ik}$,
where $\Theta$ is the Heaviside step function.
The driving force ${\bf f}_{d}=f_d{\bf {\hat x}}$ is a constant unidirectional
force applied equally to all colloids.
We initialize the system using simulated annealing in order to eliminate
undesirable transient effects
due to relaxation, and then apply the driving force. 
The equations
of motion are then integrated by velocity Verlet method for $10^{5}-10^{7}$
simulation time steps, depending on $N_{c}$. The time step $dt=0.002$.
We compute the longitudinal and transverse diffusivities $D_\alpha$ with
$\alpha=x,y$ by fitting 
$\langle [({\bf r}_i(t+\Delta t)-{\bf r}_i(t)) \cdot {\bf \hat \alpha}]^2 \rangle
=2D_\alpha \Delta t$.

The injected power 
computed for a single colloid $i$ over a time period of length $\tau$ is
given by: \begin{equation}
J_{\tau}=\int_{t}^{t+\tau}{\bf f}_{d}\cdot{\bf v}_{i}(s) ds
\end{equation}
where ${\bf v}_{i}$ represents the instantaneous velocity of colloid $i$. 
A particle which moves opposite to the direction of the driving force
makes a negative contribution to the entropy.
We measure $J_\tau$ for a series of individual particles in a single run
and combine this data to obtain $p(J_\tau)$.
We identify $J_\tau$
for a variety of $\tau$
ranging from a minimum of 10 simulation time steps to roughly one tenth the
duration of the entire simulation.

\begin{figure}
\includegraphics[width=3.5in]{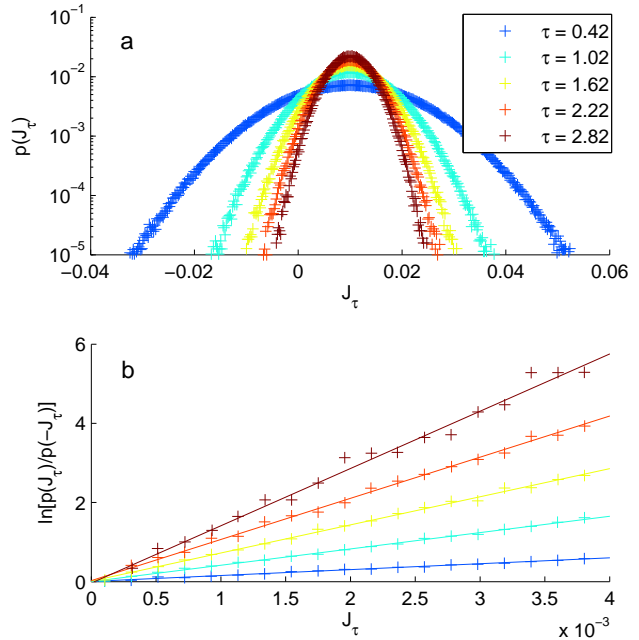} 
\caption{Demonstration of the 
FT for driven thermal particles 
in the absence of pinning at $\rho=0.5$, $T=3.0$, and 
$f_{d} = 0.1$.  (a) Probability
density function $p(J_\tau)$ of injected power for all 
observed trajectories with $\tau=0.42$, 1.02, 1.62, 2.22, and 2.82
(from center bottom to center top). (b) Fit to 
Eq.~\ref{eq:Schmickfluct}
for $\tau=0.42$, 1.02, 1.62, 2.22, and 2.82, from bottom to top. 
Linearity at various $\tau$ indicates agreement
with the 
FT.}
\label{fig:powerprobtemp} 
\end{figure}

We first consider a system at $f_d=0.1$ with no pinning but with finite
thermal fluctuations $T=3.0$
at a colloidal density of $\rho = 0.5$. 
Figure~\ref{fig:powerprobtemp}(a) shows $p(J_\tau)$ for $\tau=0.42$, 
$1.02$, $1.62$, $2.22$, and $2.82$.
Over this range, $p(J_\tau)$ is normally distributed with a slight rightward
skew due to the applied drive, and for
increasing $\tau$ the distribution sharpens. 
Equation \ref{eq:Schmickfluct} is certain to be followed since
$p(J_{\tau})=C\exp(-\left(\langle J\rangle-J_{\tau}\right)^{2}/2\sigma^{2})$
and hence $\log\left(p(J_{\tau})/p(-J_{\tau})\right)
=(-1/2\sigma^{2})[(\langle J\rangle-J_{\tau})^{2}-(\langle J\rangle+J_{\tau})^{2}]
\varpropto J_{\tau}$.
The validity of Eq.~1 for this system is
illustrated by the linear fits in Fig.~\ref{fig:powerprobtemp}(b). 

\begin{figure}
\includegraphics[width=3.5in]{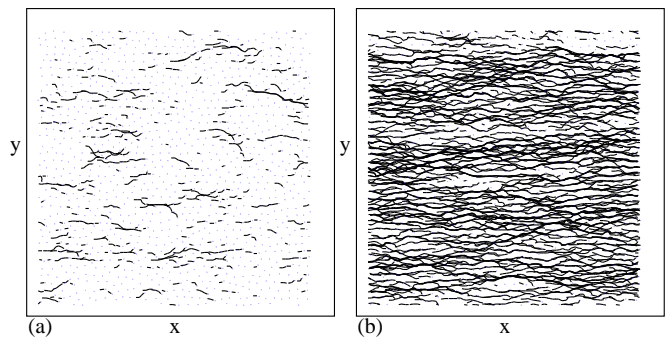} 
\caption{Colloid positions (dots) and trajectories (lines) during
40000 simulation time steps at $\rho=0.5$, $A_{p}=0.5$, and  
(a) $f_{d}=0.27$; (b) $f_d=0.34$. 
}
\label{fig:diagram} 
\end{figure}

\begin{figure}
\includegraphics[width=3.5in]{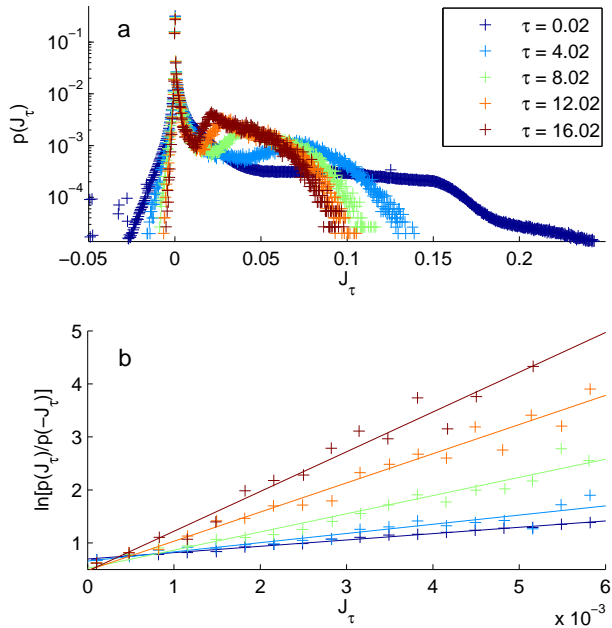} 
\caption{Demonstration of FT in a 
nonthermal system with quenched
disorder. (a) 
$p(J_\tau)$
for all observed trajectories at $\rho=0.5$ with $A_{p}=0.5$ and
$f_{d}=0.34$ at
$\tau=0.02$, 4.02, 8.02, 12.02, and 16.02 (from upper right to lower right).
(b) Fit to Eq.~\ref{eq:Schmickfluct} 
showing agreement with the FT.
Bottom to top: $\tau=0.02$, 4.02, 8.02, 12.02, and 16.02.}
\label{fig:powerprob} 
\end{figure}

We next repeat the procedure used to obtain Fig.~1 in a system with $f_d=0.34$,
no thermal fluctuations ($T=0$), and which contains pinning sites with
$A_p=0.5$.
For $f_{d}<0.33$ the particles are pinned and there
is no nontransient motion, as illustrated in 
Fig.~\ref{fig:diagram}(a).   Just above the depinning
transition at $f_{d} = 0.34$,  the particle motion persists
with time and the trajectories are highly disordered as shown in 
Fig.~\ref{fig:diagram}(b).  
Approximately one third of the colloids are
pinned at any given time; however, all of the 
particles take part in the motion.
In Fig.~\ref{fig:powerprob}(a) we plot the strongly non-Gaussian 
$p(J_\tau)$ that appear in the absence of thermalization
for $\tau=0.02$, 4.02, 8.02, 12.02, and 16.02 
at $f_{d} = 0.34$.  
The $\tau=0.02$ curve,
most closely representative of the instantaneous distribution, peaks
at $J_{\tau}=0$ and is skewed in the positive direction by the applied
drive.   
Despite the strong non-Gaussianity of the PDF's, the ratio of entropy-producing
to entropy-consuming trajectories 
is in agreement with the fluctuation theorem of Eq.~\ref{eq:Schmickfluct},
as shown in Fig. \ref{fig:powerprob}(b). 

The quality of the fit to Eq. \ref{eq:Schmickfluct} 
for fixed $\rho$ and $A_{p}$ depends on $f_d$.
It is known from earlier studies 
that systems with depinning transitions can exhibit a number
of different dynamical regimes as a function of external drive, 
including a completely pinned phase where there is no motion, a stable
filamentary channel phase just at depinning where a small number of
particles move in periodic orbits \cite{Vortex1}, 
chaotic flow at higher drives when 
the filaments change rapidly with time
\cite{Vortex1,Vortex2}, and a dynamically recrystallized phase at even higher
drives where the particle paths are mostly ordered
\cite{Vortex2}. 
To quantify the quality of the fits to Eq.~1 
we calculate the Pearson product-moment
correlation coefficient $r$ \cite{Pearson}, which is a measure of 
the linearity of the relation between two variables. 
In Fig.~4(a) we plot the mean dissipation
$\eta \langle v\rangle/f_{d}$ versus $f_d$ for the system in Fig.~2(b), along
with the corresponding longitudinal and transverse diffusivities
$D_x$ and $D_y$. 
In Fig.~4(b) 
we show the value of $r$ for varied $f_{d}$
and for all $\tau<\tau_{c}(f_{d})$, where 
$\tau_{c}(f_d)=\sup\left\{t \mid r(f_{d},\tau) \ge 0.5 \ \forall \ \tau < t\right\}$.
Agreement with the FT, indicated by $r\approx 1$, 
holds over the largest range of $\tau$ at
$f_d\approx 0.34$, near the depinning threshold and coinciding with peaks
in both $D_x$ and $D_y$.  Here, the colloids flow in plastic fluctuating
channels, as shown in Fig.~2(b).
At lower drives $f_d<0.33$, $D_x$ and $D_y$ are much smaller, the motion
in the system is very unstable, and the particles
flow only through short-lived filaments, as
shown in Fig.~2(a).
We are unable to determine whether the FT fails to hold for
$f_d<0.33$ since our measurement in this regime is dominated by rare events
and our statistics remain poor over computationally accessible time scales.
For higher drives $f_d \gtrsim 0.4$,
the colloids begin 
to form 
an ordered crystal structure 
similar to that found in vortex systems at sufficiently high driving
\cite{Vinokur,Vortex2},  
and both $D_x$ and $D_y$ drop.
The FT continues to hold for small $\tau$ at increasing $f_d$, with the
maximum value of $\tau$ at which $r>0.5$ decreasing with increasing
$f_d$.
On short time scales, the particles experience a ``shaking
temperature'' $T_s$ which decreases as 
$T_{s} \propto 1/f_{d}$  \cite{Vinokur}.
Due to particle-particle interactions, on longer time scales the
particles are effectively caged in a co-moving reference frame and no longer
undergo long time diffusion.  As a result, the FT fails to hold on
longer time scales.

\begin{figure}
\includegraphics[width=3.5in]{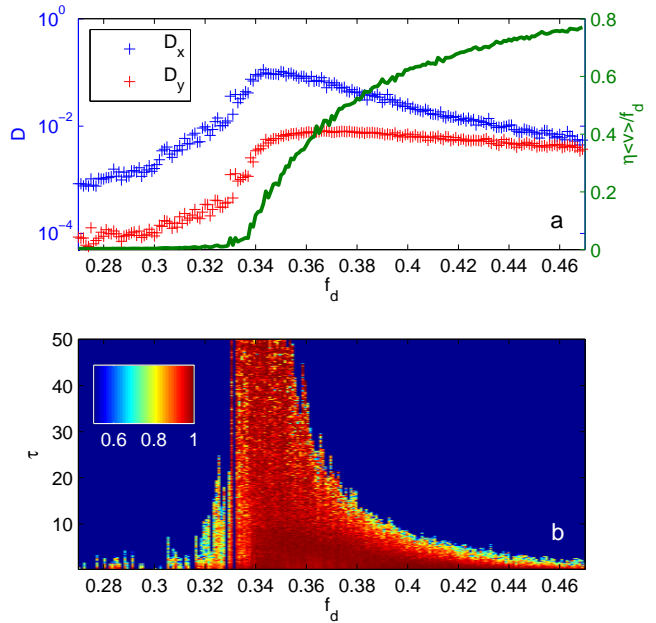} 
\caption{Limits of regime in which 
FT is verified 
in a system with $\rho=0.5$ and $f_p=0.5$.
(a) Solid curve: mean dissipation $\eta\langle v\rangle/f_{d}$ vs $f_d$,
relating colloid displacements to applied
drive. 
Upper crosses: longitudinal diffusivity $D_x$ vs $f_d$.  Lower crosses:
transverse diffusivity $D_y$ vs $f_d$.
(b) Pearson product-moment correlation coefficient 
$r$ of the fit 
$\log(p(J_{\tau})/p(-J_{\tau}))=mJ_{\tau}+b$
as a function of $\tau$ and $f_d$. 
Values closer to $1$ indicate better agreement with the FT.
The FT holds over the largest range
of $\tau$ in the fluctuating plastic flow regime near $f_d\approx 0.34$
illustrated in Fig.~2(b).  
}
\label{fig:fitqual} 
\end{figure}

\begin{figure}
\includegraphics[width=3.5in]{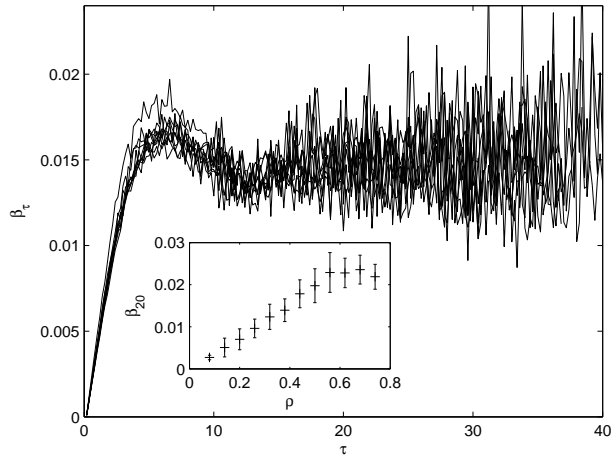} 
\caption{Effective temperature 
$\beta_\tau$ in a nonthermal system with quenched
disorder at $\rho=0.5$, $A_p=0.5$, and $f_d=0.345$. We overlay ten curves,
each representing one realization with a unique random seed.
The asymptotic value $\beta_{\infty}\approx 0.015$.
Inset: $\beta_{20}=\langle \beta_{15<\tau<25}\rangle $ calculated over a range
of $\rho$.}

\label{fig:efftempsamp} 
\end{figure}

As described previously, the FT allows the definition
of an {}``effective temperature'' $\beta_{\tau\rightarrow\infty}$ when sufficient entropy-destroying
trajectories of duration exceeding the microscopic time scales of
the system can be sampled. This necessarily involves a balance of
time scales 
since the second law of thermodynamics guarantees that $p(J_{\tau}<0)=0$
as $\tau\rightarrow\infty$.
In Fig.~\ref{fig:efftempsamp}, we plot $\beta_{\tau}$ versus $\tau$
showing the existence of an asymptotic 
effective temperature $\beta_{\infty}$ in 
a nonthermal system with quenched
disorder.  When we vary the initial configurations of the
particle positions by changing the random simulation seed,
we consistently find an asymptotic value
of $\beta_\infty \approx 0.015$ for $\tau \gtrsim 15$. Equivalently, 
this indicates that
the slope of our fits obtained as in Fig.~\ref{fig:powerprob}(b)
scales such that $\tau/S_{\tau}$ reaches a constant value at large
$\tau$. 

We 
do not observe 
significant variation in $\beta_{\infty}$
with $f_{d}$; however, as noted previously, 
we can only define an effective temperature for those values of $f_d$ where
the FT holds over a wide range of $\tau$, which limits us to drives near
the depinning threshold where plastic flow occurs.
To compare $\beta_{\infty}$ across
different $\rho$, we 
perform our measurement at $f_d=1.03f_c$ for each $\rho$, where $f_c$ is
the depinning force at that value of $\rho$.
This places us within the plastic flow regime for every $\rho$ considered
here.
The inset of Fig.~\ref{fig:efftempsamp} indicates that
there is
an apparently linear increase in $\beta_{\infty}$ with increasing
particle density saturating at 
$\rho\approx 0.6$.

Experimentally testing the FT theorem for systems of collectively 
interacting particles in the presence of quenched disorder 
could be done in several ways.
For superconducting vortices, the particle trajectories 
could be measured directly
using various imaging techniques \cite{Kes}. 
Recently it has been demonstrated that 
a single vortex can be dragged through a sample \cite{Moler}, 
making it possible to perform a vortex experiment analogous to the colloid
experiment of Wang {\it et al} \cite{WangSevick}.
The most straightforward measurement would be to monitor
the voltage fluctuations \cite{Higgins} 
at a constant drive to measure the power dissipation. 
A similar approach could be used 
to study conduction fluctuations in Wigner crystal
systems \cite{Cooper}. 

We have shown that the fluctuation theorem of 
Gallavotti and Cohen
can be applied to a nonthermal system
of collectively interacting particles driven over random disorder in
the nonlinear regime above the depinning threshold where the particles
flow plastically.
This result indicates that the fluctuation theorem may be generalized  
to a wide class of systems exhibiting fluctuating
dynamics near a transition from pinned to moving, including
magnetic domain walls, vortices in type-II superconductors,
and sliding Wigner crystals.  
It would also be interesting to apply this approach 
to analyze other non-thermal systems 
that exhibit similar crackling noise such
as dislocation dynamics.        

This work was carried out under the auspices of the NNSA of the U.S. DOE at
LANL under Contract No. DE-AC52-06NA25396. J.A.D. was supported by the
Krell Institute Computational Science Graduate Fellowship, U.S. DOE grant
DE-FG02-97ER25308.

\end{document}